# Quantum kernel learning Model constructed with small data

**Takao Tomono**[1, 2,3]\*  and **Kazuya Tsujimura**[4]

[1]*Graduate school of Media and Governance, Keio Univ., Fujisawa, Kanagawa 252-0882 Japan*

[2]*Graduate school of Science and Technology, Keio Univ., Yokohama, Kanagawa, 223-0061 Japan*

[3]*Keio Univ. Sustainable Quantum Artificial Intelligence Center (KSQAIC), Keio Univ., Minato, Tokyo, 108-8345 Japan*

[4]*Digital Innovation Division, Toppan Holdings, Tokyo, 110-8560, Japan*

e-mail address: takao.tomono@ieee.org

**ABSTRACT**

We aim to use quantum machine learning to detect various anomalies in image inspection by using small size data. Assuming the possibility that the expressive power of the quantum kernel space is superior to that of the classical kernel space, we are studying a quantum machine learning model. Through trials of image inspection processes not only for factory products but also for products including agricultural products, the importance of trials on real data is recognized. In this study, training was carried out on SVMs embedded with various quantum kernels on a small number of agricultural product image data sets collected in the company. The quantum kernels prepared in this study consisted of a smaller number of rotating gates and control gates. The F1 scores for each quantum kernel showed a significant effect of using CNOT gates. After confirming the results with a quantum simulator, the usefulness of the quantum kernels was confirmed on a quantum computer. Learning with SVMs embedded with specific quantum kernels showed significantly higher values of the AUC compared to classical kernels. The reason for the lack of learning in quantum kernels is considered to be due to kernel concentration or exponential concentration similar to the Baren plateau. The reason why the F1 score does not increase as the number of features increases is suggested to be due to exponential concentration, while at the same time it is possible that only certain features have discriminative ability. Furthermore, it is suggested that controlled Toffoli gate may be a promising quantum kernel component.

**INTRODUCTION**

In recent years, quantum computers that perform calculations using the principles of quantum mechanics have been attracting attention. The development of quantum computers based on various principles is accelerating. Such quantum computers are expected to be used for the following problems: 1) quantum simulation[1–3], 2) quantum cryptography[4–6], 3) mathematical optimization[7–9], and 4) machine learning[10–15]. Machine learning is calculated using linear algebra and matrices. Since quantum computing involves linear algebra and matrix calculations, it has the advantage of being compatible with classical machine learning. Therefore, quantum machine learning is expected to have great advantages, and we are currently conducting research for quantum machine learning.

In machine learning classification, the key is how to form a separation surface in the quantum Hilbert space[15,16]. The expressive power of the classification space is very important to form a complex separation surface. The expressive power of the quantum kernel space may be superior to that of the classical kernel space[13,16]. Based on this idea, we would like to effectively utilize the quantum kernel space to solve social issues. One of the social issues is image inspection using machine learning. Image recognition algorithms for such inspections include CNN (convolutional neural network)[17–19], VAE (variational autoencoder)[20–22], GAN (generative adversarial network)[23–25], logistic regression[26–28], random forest[29–31], boosting[32–34], SVM (support vector machine)[35–37] and so on. CNN, VAE, and GAN need expensive computational costs because they use GPUs. Compared to logistic regression, random forest, and boosting, SVM is known to be able to build learning models with less data. Moreover, we can create a complex separation surface by using the kernel trick.[38–40] In this time, we would like to make full use of the expressive power of the quantum kernel space by using the quantum kernel trick[38,41].

Now, in image inspection, anomaly detection is a very important technology. Industrial products are standardized, and image inspection is simple, but it is not easy to detect various anomalies. In addition, abnormal images are not uniform and there are different types. There have been several reports on the potential of quantum machine learning and quantum kernel estimation to perform image classification.

In our previous work[41–43], we demonstrated high performance (accuracy and F1 score) at the learning model construction stage using small datasets compared to classical machine learning. We conducted experiments by partially applying our learning model to image inspection at factory production sites. As a result, we found that our quantum machine learning produced higher evaluation indices than classical machine learning. Since the data of industrial products at the factory cannot be made public, a trial of constructing a learning model using quantum machine learning was conducted using apples as agricultural products with similar shape characteristics.

Industrial products are quality controlled and numerical data have performance within 3σ. As agricultural products are not standardized, complex separation surfaces are required to construct a learning model, unlike industrial products.

Our goal is to use quantum kernel tricks to build a system that can identify product and equipment anomalies without large amounts of data and with small amounts of data. Products include not only industrial products but also agricultural products. Products must not be shipped with abnormalities. Therefore, we believe that the quantum advantage in this field is the ability to identify abnormalities and are conducting research in this area.

In this time, we prepared 11 types of quantum kernel to investigate the influence of each two qubits gate. We use apple with internal vine cracks as an anomaly, and we created our own new dataset. Using the dataset, we obtained image information through pre-processing and feature values (attributes) from principal component analysis. Promising quantum kernels were selected based on the relationship between the calculated features (number of principal components) and the evaluation index using a quantum simulator. After screening with the quantum simulator, we confirmed the value of the evaluation index with a quantum computer. We discuss the effects of the controlled rotation gate and the CNOT gate in the quantum kernel we prepared. Furthermore, we discuss the impact on the performance indicators as the number of features increases. Finally, we propose a promising quantum kernel.

**SMALL DATASETS**

We explain the data sets we generated. We received 500 commercially available apples and found 33 anomaly apples. We confirmed that 7% of the apples on the market are apples with an invisible vine crack. A total of 66 normal and anomaly apples were used as a data set, which is different from previous work[41]. We have a lot of normal data (467 pieces). Therefore, we randomly obtained normal data from the 467 pieces and selected 24 pieces as training data and 9 pieces as test data. The number of anomaly data is limited to 33 pieces. We randomly selected 24 pieces as training data and 9 pieces as test data.

The judgement of normal or invisible anomaly apples is predicted using equipment shown in Figure 1. We get these apples from the market. We take pictures after illuminating the LED from the bottom of the apple. Then, after image processing, we obtain binary images. To know the internal situation, we cut them in half with a knife. As invisible anomaly, there are apples with browning inside and apples with vine cracks. There are two types of normal apples: normal apples with nothing inside (0) and browning apples (0*). There are also two types of anomaly apples: apples with vine cracks only (1) and apples with vine cracks and browning (1*). In other words, there are four types. Each speech bubble is an enlarged view.

The resolution of the photographed image is 4032×3024 and it is necessary to detect patterns other than the vine that appear in an area of 120×80. The judgement must be made in an area of about 3% of the total image. The 4032×3024 resolution is too large to create a learning model. The learning model is built using the image size (403×302). The image is used to distinguish between normal and anomaly for training.

**QUANTUM KERNEL AND STEP OF CLASSIFICATION**

By means of a nonlinear mapping φ(x) embedding the data into the quantum feature space, it can be expressed in the feature space as follows.

$$\kappa(x_i, x_j) = \left| \left\langle \phi(x_j)^\dagger \middle| \phi(x_i) \right\rangle \right|^2 \qquad (1)$$

First, we prepare quantum state $\varphi(x) = U(0)|0\rangle$ as a data encoding from classical to quantum data. Second, we prepare $U(x_j)^\dagger U(x)|0\rangle$ as the initial state of the quantum circuit to obtain the inner product $\kappa(x_i, x_j)$. The probability of measuring on $|0\rangle$ for all qubits is as follows.

$$\langle 0|U^{\dagger}(x_i)U(x_j)|0\rangle\langle 0|U^{\dagger}(x_j)U(x_i)|0\rangle = \langle \phi(x_i)|\phi(x_j)\rangle\langle \phi(x_j)|\phi(x_i)\rangle \qquad (2)$$

Here $U(x)$ is the inner product between the quantum encoded data using quantum kernel estimation. Each feature map is then embedded into the inner product to optimize the parameters. The matrix component of the entire Gram-matrix is obtained from a combination of the inner product. The parameters of the kernel estimation are optimized using rotation gates with/without entanglement in Eq. (2). We use RBF as classical kernel of SVM in this work.

Figure 2 shows quantum kernel circuits diagram. Fig.2(a) shows quantum kernel circuits diagram. Fig.2(b) shows the details of $\phi(x_i)$ as equation (1) and (2).

QK0 and QK1 are circuits with only rotation gates (H and H Ry). QK2 to QK10 were prepared to create efficient quantum kernels with fewer gates than hardware efficient embedding (HEE)[44,45]. QK2 and QK3 are circuits that place controlled Ry and Rx gates between each qubit and the next qubit in a staircase pattern. QK4 is a circuit that places a controlled Ry gate between each qubit and the bottom qubit. QK5 is a circuit in which Rz is inserted between the Ry control gates of QK4. QK6 is a circuit in which CNOT gates are arranged in a staircase pattern with Ry inserted between each CNOT gate. QK7 is a circuit in which each qubit and the lowermost qubit are connected by a CNOT gate and Ry is inserted between each CNOT gate. QK8 is a circuit in which a Rz gate is inserted instead of the Ry gate in QK7. QK9 is a circuit in which Rz is placed on each qubit at the end of the quantum circuit in QK7. QK10 is a circuit in which controlled Toffoli gate, is placed in place of each CNOT in QK9. The experiment was essentially carried out using a quantum simulator (ibm_qasm_simulator). For final confirmation, we used a real quantum computer (ibmq_Osaka of the IBM Quantum Platform).

Figure 3 shows step of classification. we prepare training data and test data. Using the datasets of apple we created, we perform 1 preprocessing. Then, 2 principal component analysis is performed to extract features. Then, 3-1 classical and 3-2 quantum kernels are generated using the features, and 4 Classical SVM embedded kernel is performed to build a learning model. The learning model is used to predict the test data.

**F1 SCORE ON QUANTUM SIMULATOR**

Table 1 shows the contribution ratio (CR) and cumulative contribution ratio (CCR) obtained by principal component analysis of the image. The first to tenth principal components are obtained. Although there is no clear criterion for the cumulative contribution ratio (CCR), a rough image can be reproduced if the CCR is 0.5 or higher; a CCR of 0.85 or higher is sufficient to reproduce the image. The cumulative contribution ratio (CCR) of principal components 4, 7 and 10 are 0.662, 0.744 and 0.799 respectively. Confirming the reproducibility of the anomaly apple images, it is considered that the first, second and third principal components mainly represent the external shape, browning of the apple and internal vine crack. Here, the number of principal components corresponds to features value.

Figure 4 shows the relationship between the F1 scores for the principal components of each quantum kernel and the conventional kernel RBF. Fig. 4(a) shows the results for the classical kernel RBF and QK0-QK5. The horizontal axis is the feature corresponding to the cumulative contribution of the principal components. The vertical axis is the F1 score. If the feature value is 3, it means the cumulative contribution of the first to the third principal component. If the feature value is 7, it means the cumulative contribution of the first to the seventh principal component. First, the F1 scores of each kernel are compared when the feature value is 3, and then the trends in the F1 scores are compared as the feature size increases.

QK0 and QK1 are quantum circuits with one H-gate, one H-gate and one rotating gate Ry for each qubit. Their F1 value is 0.2 higher than the F1 score of the classical kernel RBF. However, when the feature value increased from 3 to 7, the value of the F1 score increased by 0.1.

QK2 and QK3 are quantum circuits with a staircase structure of control gates Ry and Rx. The F1 score of QK2 and QK3 are more than 0.1 larger than the F1 score of RBF, but 0.05 smaller than the F1 score of QK0. QK4 is a quantum circuit with a controlled rotation gate between each qubit and the bottom qubit. The F1 scores of QK4 and QK5 are more than 0.1 larger than the F1 score of QK0, QK1, QK2 and QK3, but 0.1 smaller than the F1 score of QK2. The F1 scores for QK2, QK3, QK4 and QK5 remained almost constant as feature value increased from 3 to 7.

From the above, it can be considered that the quantum kernel has greater discriminative power than the classical kernel when the feature value is 3. However, the reason why the F1 score remains almost constant even as the feature value increases is because the images represented by the fourth to seventh principal components have little effect on the vine crack, and therefore the value of the F1 score is considered to be unaffected.

Figure 4(b) compares the relationship of QK6 to QK10 with the classical kernel RBF. For a feature value of 3, the F1 score of QK7 and QK8 are more than 0.15 larger than the RBF. The QK6 is about 0.1 larger than QK7 and QK8. Furthermore, the F1 score of QK9 and QK10 are more than 0.3 larger than the classical RBF. The F1 score of QK6 is almost the same as that of QK1. The behavior of QK6 is similar to that of QK1.

QK6 has a staircase CNOT gate. Except for the first qubit, it has a Ry rotation gate from the second qubit onwards similar to QK1. Therefore, the effect of the CNOT gates on the F1 score was not significant in the case of QK6. For QK7, QK8, QK9 and QK10, the F1 score becomes larger as the feature value increases, with the F1 score for QK7 and QK8 increasing by 0.2 and the F1 score for QK9 and QK10 increasing by 0.1. This increase could be considered a dominant difference, but it was equal to or less than that of the classical kernel.

From the above it can be concluded that the CNOT gate is dominant except for QK6. Throughout the apple, including external shape, browning and internal vine crack, the quantum kernel is considered to have a higher discriminative capacity for anomalies. The F1 scores of QK9 and QK10 are the highest among these quantum kernels and are considered promising quantum kernel candidates.

**F1 SCORE AND AUC ON QUANTUM COMPUTER**

After obtaining the above results using the simulator, we use ibm_Osaka to confirm the behavior of true quantum computer. Figure 5 shows the ROC-AUC curve (step-shaped curve) of quantum computer compared to classical computer and quantum simulator. Here, QC and QS means quantum computer and simulator. The left figure shows the ROC-AUC curve of RBF, QS and QC on QK9, and the right figure shows the ROC-AUC curve of RBF QS and QC on QK10. As reference data, we also draw the position of random model (black dashed line) and ideal learning model (red dashed line). The dashed line from False Positive Rate (FPR)=0, True Positive Rate (TPR)=0 to FPR = 1, TPR=1 indicates the random model. Axis of FPR=0 and TPR=1 indicates ideal learning model. The AUC of classical

RBF is drawn near the random model and the numerical data is 0.62. on the other hand, for QK9, the behavior of the AUC curve for the quantum computer was the same as that for the quantum simulator. The numerical data of AUC at that time were both 0.90, as shown in the figure.

For QK10, the behavior of the ROC-AUC curve for the quantum computer was lower than that of the classical computer. As shown in the figure, the numerical data of AUC value at that time was 0.89 for the quantum simulator and 0.59 for the quantum computer. For QK10, the behavior and numerical data on quantum computer were significantly different from those on the quantum simulator.

To confirm the reliability of the numerical data, measurements were performed at least 3 times on the quantum computer. In addition, when measurements were performed 5 times for classical computer, it was found that the error was within approximately 2 %.

Figure 6 shows reproducibility of calculations using a quantum simulator and computer. We compare with the numerical data of AUC and the F1 score for the classical kernel RBF and the quantum kernels QK0, QK9, and QK10. The quantum kernels QK0, QK9, and QK10 were measured, and the maximum, minimum, and average values were calculated. For quantum kernels other than QK10, there appears to be little difference in the values of each evaluation index (AUC and F1-score) between the quantum simulator and the quantum computer.

To investigate the cause of the difference between the quantum simulator and the quantum computer in QK10, the circuit depth of the quantum circuit was investigated, and the results are shown in Table 2. Since there was no difference between the quantum simulator and the quantum computer in QK0, QK1, and QK9, it is believed that there is no problem with a circuit depth of up to 32, but since a problem occurred at a circuit depth of 273, it is believed that errors accumulated between 32 and 273. This confirmed that the quantum circuit depth affected the generation of errors in the quantum computer.

From the above, we confirmed that the results of the quantum simulator and the quantum computer are almost the same. We also confirmed that as the depth of the quantum circuit increases, the quantum computer becomes noisy, and the calculation becomes difficult at the circuit depth of QK10. We also found that QK9 is a promising quantum kernel.

**DISCUSSION**

Based on the experimental results, the features obtained after feature extraction by principal component analysis correspond to each qubit. As we described already above, the first principal component obtained by the principal component analysis is considered to be the external shape of the apple, the second principal component is the browning, and the third principal component is the internal vine crack. Here, we assume that overfitting does not occur due to the small size of the data set.

When feature value is 3 (the cumulative contribution up to the third principal component is integrated), the quantum kernel was larger by more than 0.3 compared to the classical kernel. As shown in Figures 4(a) and 5(b), when the feature value is 3, the difference between the quantum and classical kernels is large, indicating that the quantum kernel is larger than classical kernel. On the other hand, as the feature size increases, the difference between the F1 scores of the quantum and classical kernels becomes smaller. This means that the advantage of the quantum kernel is less as the feature value increases.

As shown in equation (1), the quantum kernel is represented by an inner product, and from the inner product of vectors, a Gram matrix is generated. The space created by this Gram matrix generates a complex separating boundary surface. $(x_i, x_j)$ are the coordinates of the Gram matrix and $\kappa(x_i, x_j)$ are the measurements of the last individual quantum bit calculated in depth. The number of matrices in the Gram matrix is determined by the number of qubits in the quantum circuit. The gate operation (unitary called U) determines the height, and $\kappa(x_i, x_j)$ create energy gradients such as peaks, valleys and plains, and the separation boundary is determined as the result. Usually, in principal component analysis, as the cumulative contribution increases (feature value increases) in classical kernel learning, the image reproducibility improves, and the discriminative power become larger.

When the feature size is 3, the F1 score with the quantum kernel is sufficiently larger than the F1 score with the classical kernel. On the other hand, as the feature size increases, the difference between the F1 score by the quantum kernel and the F1 score by the classical kernel becomes smaller. This could be caused by a phenomenon similar to Barren Plateaus (BPs)[46–51] due to exponential concentration. On the other hand, it is possible that, perhaps, machine learning with quantum kernels may provide better discriminative power with respect to certain anomalies.

Experimental results show that controlled Toffoli gates are promising gates for quantum kernel circuits. The controlled Toffoli gate is a gate that can be decomposed into Hadamard gates, phase gates and a control NOT gates. As shown in Table 2, the depth of the QK9 circuit is 32, while that of the QK10 circuit is 273. the depth of the QK10 circuit is 8.5 times larger than that of the QK9 circuit. In this study we performed the calculations on a superconducting quantum computer, but the calculations failed due to noise and other factors. The reason for this is thought to be the topology of the superconducting quantum computer, which utilizes many swap gates, which makes the depth of the circuit deeper. However, an all-coupled ion-trap quantum computer could be computed without the use of swap gates. This has the advantage that it is likely to be able to compute with reduced errors. On the other hand, the size of the Gram matrix increases in learning (training) using data, which poses a challenge to the execution of calculations. In the case of ion-trap quantum computers, the coherence time is long, so if the size of the Gram matrix can be reduced in future calculations, the control Toffoli gate will be a promising gate with high expectations.

In this application side to quantum computer, the F1 score of QK10 using controlled Toffoli gate shows a sufficiently large F1 score in the quantum simulator, so it is thought to be very promising in practical situations where error correction becomes possible.

**CONCLUSION**

We attempted to detect anomaly for apple with internal vine cracks, which are real data with high similarity to image data from our own factory. Using the quantum kernel trick, it was demonstrated that a learning model could be built to detect a single anomaly (an internal vine crack) from 24 training data. Ten different quantum kernels were used, and performance index were evaluated according to feature value. We used a quantum simulator and a computer to examine F1 scores and finally evaluate the construction of a learning model using AUC. The results showed that the F1 score was greater with the CNOT gate than with the control rotation gate. The SVM with quantum kernel has a larger F1 score and better discriminative ability than the SVM with classical kernel. Therefore, in anomaly detection, discriminative ability is considered a quantum advantage on SVM embedded quantum kernel. As the feature value increases, the increase in the F1 score of the quantum kernel was smaller than that of the classical kernel.

Within the quantum kernel, a quantum kernel circuit (QK9) with a CNOT gate connecting each qubit to the bottom qubit, a Ry rotation gate inserted between CNOTs and an Rz gate at the end of each qubit of the quantum circuit was found to be promising. If the memory cost of the Gram matrix is reduced and fault-tolerant quantum computers are developed, controlled Toffoli gates would be a promising quantum circuit.

**A LIST OF ABBREVUATIONS**

SVM: Support vector machine; GPU: Graphics Processing Unit; CNN: Convolutional neural network; VAE: Variational autoencoder; GAN: Generative adversarial network; LED: Light Emitting Diode; RBF: Radial Basis Function; F1-score: Harmonic mean of conformity and repeatability; AUC: Area Under the Curve and AUC takes values between 0 and 1, with values closer to 1 indicating higher discriminatory power; CR: Contribution Ratio; CCR: Cumulative Contribution Ratio; FPR: False Positive Rate; TPR: True Positive Rate.

**ACKNOWLEDGMENTS**

We thanks Mr. Godo to select apples and taking high resolution pictures of apples. We thank IBM to support free trial of quantum computer. This is based on results obtained from a project, JPNP23003, commissioned by the New Energy and Industrial Technology Development Organization (NEDO). Takao Tomono is supported by the Center of Innovations for Sustainable Quantum AI (JST), Grant No. JPMJPF2221.

**AUTHOR DECLARATIONS**

1) Ethical Approval and Consent to participate

    Not applicable

2) Consent for publication

    Not applicable

3) Availability of supporting data

    Raw data were generated at Toppan Holdings. Derived data supporting the findings of this study are available from the corresponding author T.T. on request.

4) Competing interests

    The authors have no conflicts of interest to declare that are relevant to the content of this article.

5) Authors' contributions

T.T. designed research. T.T. and K.T. made datasets of apple. T.T. made basic quantum software based on quantum algorithms. K.T. conducted revised quantum software and calculate indices by classical and quantum computation. Both contributed to the discussion and T.T wrote manuscript.

6) Funding

This work was supported by the New Energy and Industrial Technology Development Organization (NEDO) [Grant No. JPNP23003] and by the Center of Innovations for Sustainable Quantum AI (JST) [Grant No. JPMJPF2221].

# TABLE

Table 1. Principal Components (PC), Contribution Ratios (CR), and Cumulative Contribution Ratios (CCR) obtained by performing principal component analysis on this image. PC stands for principal components. CR stands for contribution ratios. CCR stands for cumulative contribution ratios.

| PC | 1 | 2 | 3 | 4 | 5 | 6 | 7 | 8 | 9 | 10 |
|---|---|---|---|---|---|---|---|---|---|---|
| CR | 0.4390 | 0.1210 | 0.0593 | 0.0423 | 0.0337 | 0.0278 | 0.0210 | 0.0207 | 0.0176 | 0.0168 |
| CCR | 0.4390 | 0.5600 | 0.6193 | 0.6616 | 0.6953 | 0.7231 | 0.7441 | 0.7648 | 0.7824 | 0.7992 |

Table 2. The depth of circuits for QK0, QK1, QK9, QK10 and RBF. We use ibmq_Osaka as quantum computer when feature value is 4.

| Quantum Kernel | QK0 | QK1 | QK9 | QK10 |
|---|---|---|---|---|
| Depth of circuits | 4 | 4 | 32 | 273 |

# FIGURE

Figure 1. The process of acquiring datasets (binary images) after illumination of LED light. After visual inspection, we obtain these apples. We take pictures after illumination of LEDs from bottom of apple using equipment. Then, we acquire binary images after image processing. To know the internal situation, we cut them in half with a knife. As invisible anomaly apple, there are apples with browning inside and apples with vine cracks. The green and blue circles show a part of browning and vine cracks. There are two types of normal apples: normal apples with nothing inside (0), and browning apples (0*). The sign of * shows browning. There are also two types of anomaly apples: apples with only vine cracks (1), and apples with vine cracks and browning (1*). That is, there are a total of four types. Each speech bubble is an enlarged view.

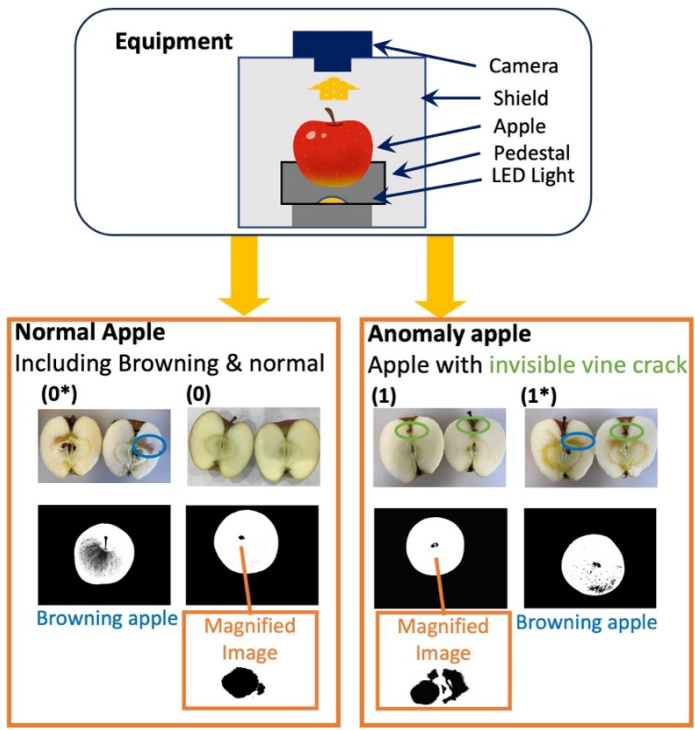

Figure 2. Quantum circuit diagram. Fig. 2(a): circuits for quantum kernel. Fig. 2(b): Circuits diagram on each quantum kernel $U_k(x_i)$. QK0 and QK1 are circuits with only rotation gates (H and H Ry). QK2 and QK3 are circuits with control rotation gates CRy and CRx arranged in a staircase between each qubit and the next qubit. QK4 is a circuit with each qubit and the bottom qubit. QK5 is a circuit with a control Ry gate between each control Ry gate of QK4. QK6 is a circuit with a CRy gate of QK2 replaced by a CNOT gate and Ry inserted between each CNOT. QK7 is a circuit with a CRy gate of QK5 replaced by a CNOT gate and Rz is replaced by Ry. QK8 is a circuit where the CRy gate of QK5 is replaced by a CNOT gate. QK9 is a circuit where the Rz rotation gate is placed at the end of the QK7 circuit and QK10 is a circuit where controlled Toffoli is used instead of a CNOT.

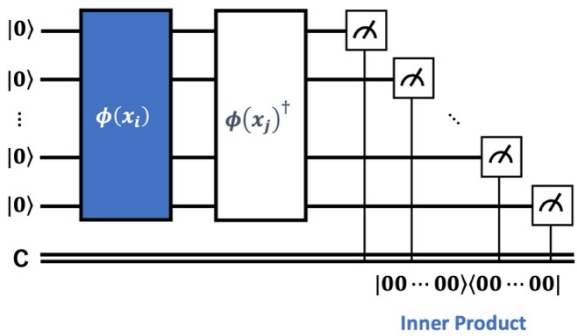

(a) Circuits for Quantum kernel

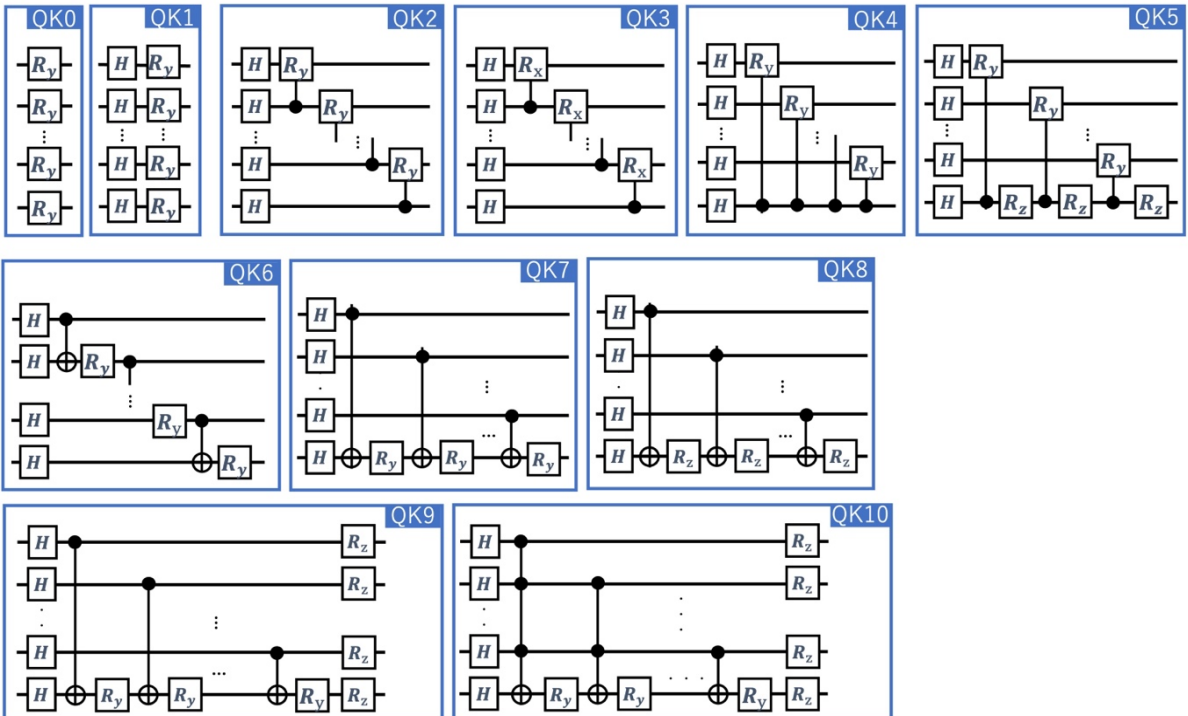

(b) Circuits diagram on each quantum kernel $U_k(x_i)$.

Figure 3. step of classification. we prepare training data and test data. Using the datasets of apple we created, we perform 1 preprocessing. Then, 2 principal component analysis is performed to extract features. Then, 3-1 classical and 3-2 quantum kernels are generated using the features, and 4 SVM is embedded to construct a learning model. The constructed learning model is used to predict the test data.

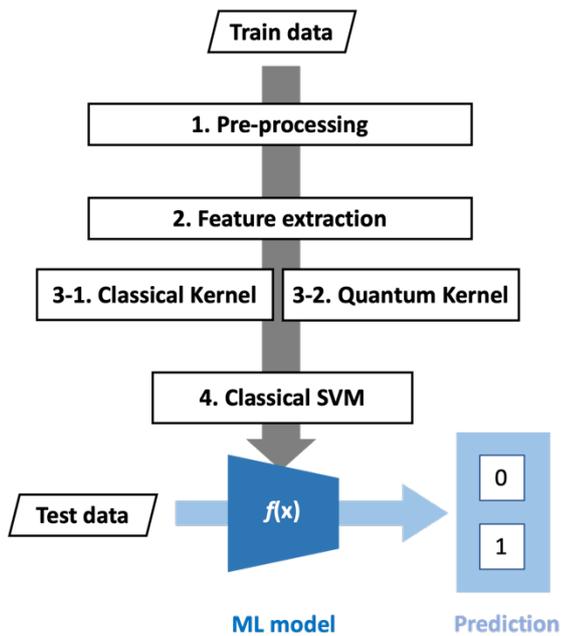

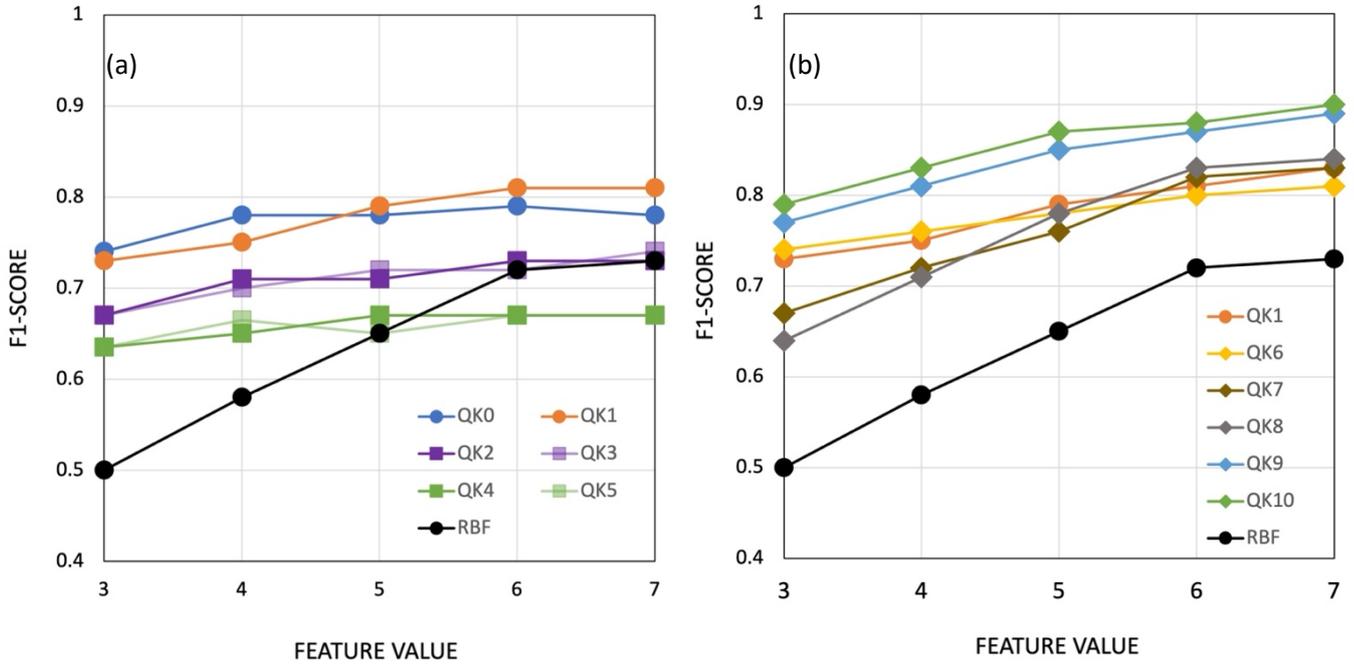

Figure 4. The relationship between Features and F1-score for each quantum kernel compared to classical kernel RBF. Fig. 4(a): the relation from QK0 to QK5 compared to classical kernel RBF. Fig. 4(b): the relation from QK6 to QK10 compared to classical kernel RBF.

Figure.5 ROC-AUC curve. The horizontal axis is the False Positive Rate (FPR), and the vertical axis is the True Positive Rate (TPR). The normal learning model construction process starts with learning from a random model at the position of the black dashed line (position of the black dashed line) and moves towards the ideal learning model (position of the red dashed line).

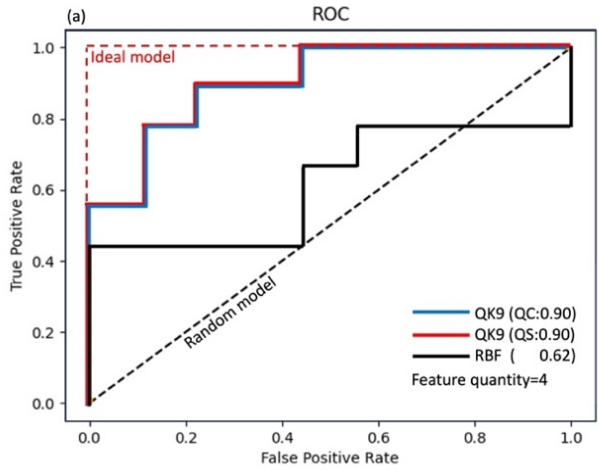
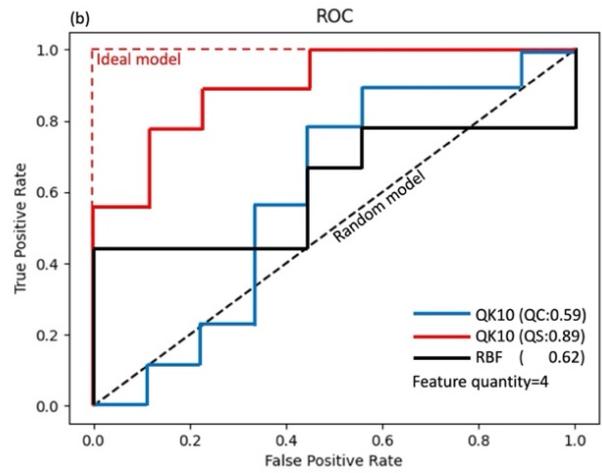

Figure.6 Reproducibility of calculations using a quantum simulator and computer. The average, maximum, and minimum values of 3-5 measurements are indicated. * represents the results of calculations using a quantum computer.

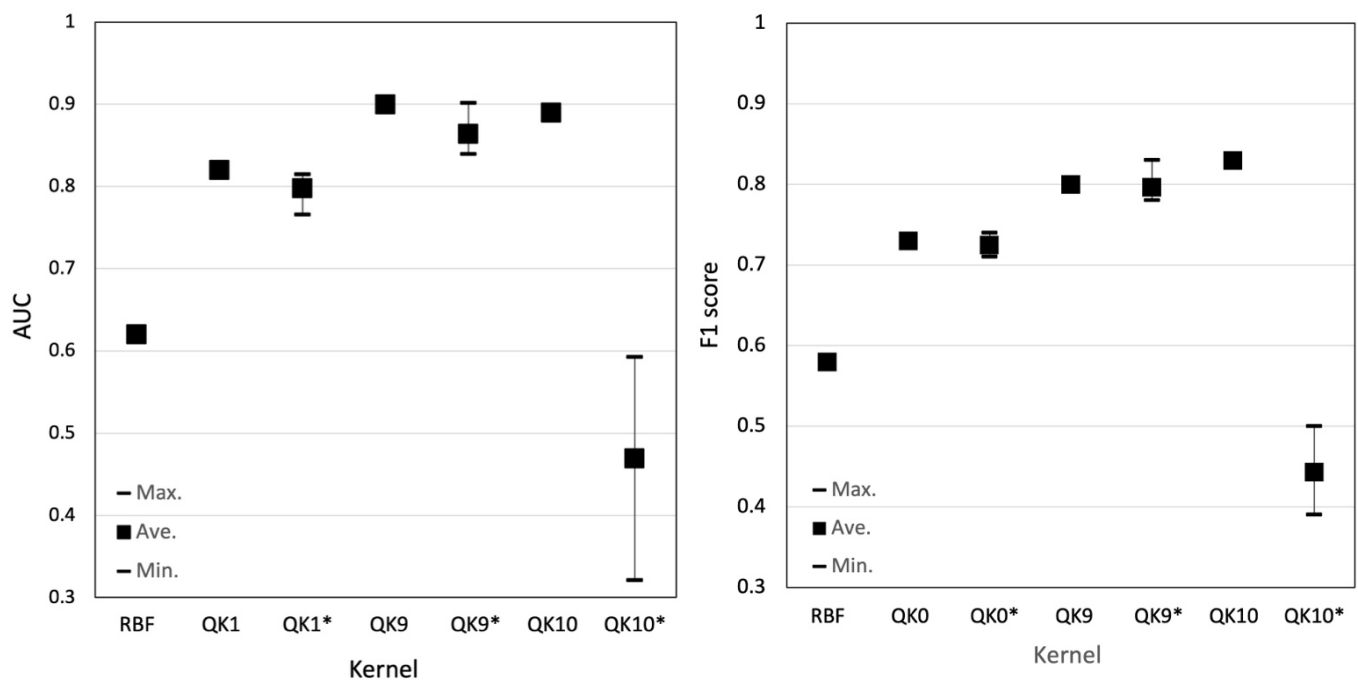